\documentclass[aps,prd,notitlepage,reprint,onecolumn,showpacs,superscriptaddress,nofootinbib,floatfix]{revtex4-1}

\usepackage{amsmath,amssymb,graphicx}
\usepackage{mathrsfs}
\usepackage{bbm}
\usepackage{bbold}
\usepackage{color}
\usepackage[colorlinks=true,citecolor=blue]{hyperref}
\usepackage{slashed}

\newcommand{\be}{\begin{equation}}
\newcommand{\ee}{\end{equation}}
\newcommand{\bi}{\begin{itemize}}
\newcommand{\ei}{\end{itemize}}
\newcommand{\bea}{\begin{eqnarray}}
\newcommand{\eea}{\end{eqnarray}}

\definecolor{bc}{rgb}{0, 0.7, 0.0}

\newcommand{\ud}{\mathrm{d}}
\newcommand{\LCm}{{\scriptscriptstyle -}} 
\newcommand{\LCp}{{\scriptscriptstyle +}}

\newcommand{\LCperp}{{\scriptscriptstyle \perp}}

\newcommand{\bracket}[2]{\bra{#1}\,#2\rangle} 
\newcommand{\bra}[1]{\langle\,#1\,|}          
\newcommand{\ket}[1]{|\,#1\,\rangle}          

\newcommand{\pathD}{\!\mathcal{D}}

\usepackage[T1]{fontenc} 

\begin{document}

\title{Back-reaction on background fields: a coherent state approach}

\author{Anton Ilderton}
\email{anton.ilderton@plymouth.ac.uk}
\affiliation{Centre for Mathematical Sciences, University of Plymouth, PL4 8AA, UK}

\author{Daniel Seipt}
\email{d.seipt@lancaster.ac.uk}
\affiliation{Lancaster University, Physics Department, Lancaster LA1
4YW, UK}
\affiliation{The Cockcroft Institute, Daresbury Laboratory, Warrington
WA4 4AD, UK}

\begin{abstract}
There are many situations in which a strong electromagnetic field may be approximated as a fixed background. Going beyond this approximation, i.e.~accounting for the back-reaction of quantum process on the field, is however challenging. Here we develop an approach to this problem which is a straightforward extension of background field methods. The approach follows from the observation that scattering in an on-shell background is equivalent to scattering between coherent states; we show that by deforming these states one can model back-reaction. Focussing on intense laser-matter interactions, we provide examples which model beam depletion and, furthermore, introduce an extremisation principle with which to determine the level of depletion in a given scattering process.
\end{abstract}

\maketitle
\section{Introduction}
There are many physical phenomena which are naturally described within the framework of ``quantum field theory in background fields''. For example, the physics of QCD in strong background magnetic fields~\cite{Kharzeev:2013jha}, pair production in the fields of heavy-ion collisions~\cite{Baur:2007fv}, extreme magnetic fields in astrophysics~\cite{Harding:2006qn} and QED in strong laser fields~\cite{Dunne:2008kc,DiPiazza:2011tq,King:2015tba} have all received a great deal of attention in the last decade.

While the mathematical structure of background field theories can be very rich~\cite{Miller:2013,Marchesiello:2015,Heinzl:2017blq}, there are inevitably physical regimes where the assumption of a background field breaks down, forcing us to go beyond this approximation. For example the back-reaction of produced particles on pair-creating fields can be significant~\cite{Kasper:2014uaa}, and it becomes important to account for beam depletion in laser-particle collisions at high laser field strengths and for dense particle bunches~\cite{Seipt:2016fyu}. The breakdown of the background field approximation may be signalled by the breakdown of the perturbative expansion of the theory, see~\cite{Fedotov:2017pho} for a review and references in the context of laser-matter interactions.  For perturbations beyond background fields in heavy-ion collisions see e.g.~\cite{Epelbaum:2013waa}. 

Going beyond the background field approximation is in general a challenging problem. One can use classical-statistical approximations and lattice simulations~\cite{Hebenstreit:2013qxa}, as has been applied to the screening of pair-producing fields by the created particles~\cite{Gelis:2013oca}. Again in the context of beam depletion and laser-matter interactions, another approach is based on approximating the gauge field to a single mode (the laser mode)~\cite{Bergou:1980cm,Bergou:1980cp,Bagrov:1990xp}, giving an exactly solvable field theory~\cite{OSS} in which to examine transitions between different states describing the laser.

Here we suggest an approach to back-reaction in laser-matter interactions which is based on the observation that scattering in a background is (as will be reviewed below) equivalent to scattering between the same initial and final coherent states, with these states describing a prescribed field~\cite{Kibble:1965zza,Frantz}.  As a step toward modelling back-reaction we therefore consider here scattering between different initial and final coherent states; we will see that this can indeed be used to model, for example, depletion of a laser field.

One motivation for our approach is that at high intensities particles emit large numbers of photons which may be better described as coherent states rather than discrete number states~\cite{Zwanziger:1973if}, hence the difference between initial and final scattering states would be at least partially coherent. A second motivation is that the coherent state origin of a background field is seldom exploited in calculations, even though it is essential for the physics -- the quantum state describing a laser cannot be a number of state of photons, for example, as such states give zero expectation value for the electromagnetic fields.  Finally, and as we will see, our approach has the advantage of being technically simple; it requires only a minor extension of the usual background field formalism.

This paper is organised as follows. In Sect.~\ref{SECT:review} we review the calculation of scattering amplitudes in background fields and the connection to coherent states. In Sect.~\ref{SECT:LSZ} we set up our depletion problem and derive reduction formulae for scattering between different coherent states. Simple applications of the formalism are outlined in Sect.~\ref{SECT:EXEMPEL}; we consider strong depletion of a weak field, weak depletion of a strong field, and strong depletion of a strong field. In Sect.~\ref{SECT:MAXIMISERING} we introduce an extremisation principle in order to identify the `most likely' level of beam depletion and give a first investigation for the process of pair creation. We work with scalar field theories throughout in order to simplify the presentation. The extension to QED is given in Sect.~\ref{SECT:CONS} along with conclusions.

\section{A coherent state model of depletion}
We begin by briefly reviewing the background field formalism and the connection to coherent states.
\subsection{Review of QED in background fields}\label{SECT:review}
\begin{figure}[t!]
\includegraphics[width=9cm]{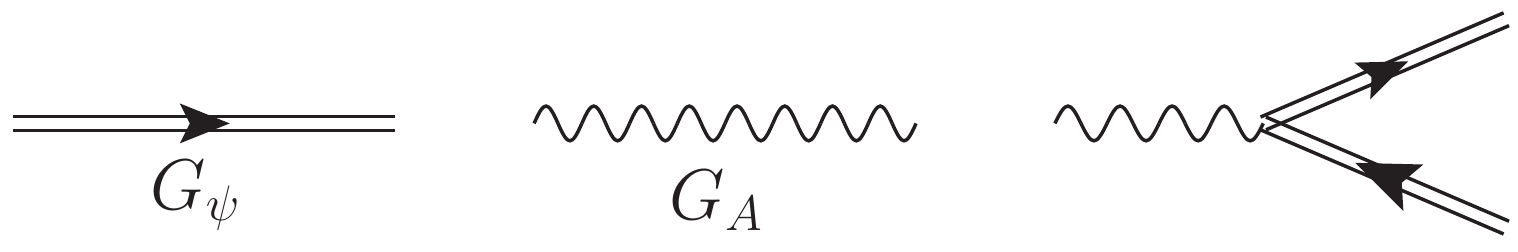}
\caption{\label{FIG:FURRY} Feynman rules for the Furry picture expansion of correlation functions, in which the interaction between matter and the background field is, in principle, treated exactly. The double line is the background-dressed fermion propagator (\ref{GPSI}).
}
\end{figure}
The action of QED with an additional background field described by the classical potential $A^\text{ext}$ is
\be\label{NYTT-S}
	S = \int\!\ud^4x \ \bar{\psi} \big(i\slashed{\partial} - m \big) \psi - \frac{1}{4}F_{\mu\nu} F^{\mu\nu} - e \bar\psi \big(\slashed{A} +\slashed{A}^\text{ext}\big)\psi +\text{ gauge fixing + counterterms}\;,
\ee
in which the interaction terms are shifted by $A_\mu^\text{ext}$. When the coupling $eA^\text{ext}/m$ is large, $S$-matrix elements corresponding to this action are ideally calculated using the `Furry picture' expansion which (in principle) allows one to treat the strong coupling to the background $A_\mu^\text{ext}$ without recourse to perturbation theory. In this expansion the interaction with the quantised gauge field is treated in perturbation theory as normal, while the coupling to $A_\mu^\text{ext}$ is included in the `free' part of the action, such that the fermion propagator becomes the background-field dressed propagator $G_\psi$,
\be\label{GPSI}
	G_\psi = \big(i\slashed{\partial} - e\slashed{A}^\text{ext} -m \big)^{-1} \;,
\ee
and the Feynman rules are as shown in Fig.~\ref{FIG:FURRY}, where the double line represents $G_\psi$.
This propagator, and the corresponding asymptotic wavefunctions following from LSZ reduction, are what is used customarily in strong-field QED calculations, 
see~\cite{Mitter:ActaPhysAustr1975,Ritus:JSLR1985,Ehlotzky:ProgPhys2009,
Dunne:2008kc,DiPiazza:2011tq,Burton:2014wsa,King:2015tba,Seipt:2017ckc}
for reviews, and~\cite{Loetstedt:PRL2007,Bulanov:PRL2010,Mackenroth:PRL2010,Heinzl:2010vg,Krajewska:PRA2011,Seipt:PRD2012,Gies:2013yxa,Ilderton:PLB2013,Blackburn:PRL2014,Seipt:PRA2014,King:PRA2015,Heinzl:PRD2016,Seipt:NJP2016, Dinu:PRL2016, Meuren:PRD2016, Hartin:2016sha,Harvey:PRL2017} for a selection of recent papers.

Now let us make the connection to coherent states. Let the \textit{S}-matrix be, highlighting only the dependency on the gauge potential, $\mathcal{S}[A+A^\text{ext}]$. We are then interested in the \textit{S}-matrix elements
\be
	\bra{\text{out}} \mathcal{S}[A+A^\text{ext}] \ket{\text{in}} \;,
\ee
in which `in' and `out' represent injected, scattered or produced electrons, positrons and photons. We assume now that the background field is a solution of Maxwell's equations in vacuum, $\partial_\mu F^{\mu\nu}_\mathrm{ext}=0$, which is a standard assumption for laser-matter interactions.
Because the background is on-shell, it has the same Fourier modes as the quantum field, and hence the sum $A + A^\text{ext}$ is, in Fourier space,  just a sum of translated operators $\sim a(\mathbf k) + z(\mathbf k)$ in which $z(\mathbf k)$ is the (appropriately normalised as given below) positive energy Fourier mode of $A_\mu^\mathrm{ext}$.
We can therefore extract the background from the $S$-matrix using the standard displacement, or translation, operator
$$D(z) = \exp \bigg( \! \int \! \ud^3 \mathbf k \: a ^\dagger({\bf k}) z({\bf k}) - a({\bf k}) \bar{z}({\bf k}) \bigg)\,,$$
(suppressing polarisation labels for the moment) which obeys the commutation relations
\be\label{DaD}
	D^\dagger(z) a({\bf k}) D(z) = a({\bf k}) + z({\bf k}) \;, \qquad D^\dagger(z) a^\dagger({\bf k}) D(z) = a^\dagger({\bf k}) + {\bar z}({\bf k}) \;.
\ee
Then we have the long-established result~\cite{Kibble:1965zza,Frantz}
\be\label{vanligt}
	\bra{\text{out}} \mathcal{S}[A + A_\text{ext}] \ket{\text{in}} = \bra{\text{out}} D^\dagger(z) \mathcal{S}[A] D(z)\ket{\text{in}} = \bra{\text{out};z} \mathcal{S}[A] \ket{\text{in};z} \;,
\ee
where, in the last step, we use the fact that $D(z)$ acting on the vacuum\footnote{This requires a mild assumption on the momentum support of the `in' and `out' states which we explain below.} creates a normalised coherent state $\ket{z}$. Hence scattering in an on-shell background field is equivalent to scattering between asymptotic coherent states. Note that this derivation does not require any assumptions on the form of the $S$-matrix. Furthermore, the background  field can now, using (\ref{vanligt}), be re-interpreted as the expectation value of the potential in the coherent state:
\be
	A_\mu^\text{ext}(x) = \bra{\text{out};z} A_\mu(x) \ket{\text{out};z} = \bra{\text{in};z} A_\mu(x) \ket{\text{in};z}\;.
\ee
This makes it explicit that we have no back-reaction, since the expectation value of the field, and the mean number of photons in each mode, $|z({\bf k})|^2 = \bra{z} a^\dagger (\mathbf k) a(\mathbf k)\ket{z}$, are the same in both the initial and final states.

\subsection{Beam depletion}\label{SECT:LSZ}
\begin{figure}[t!]
\includegraphics[width=7cm]{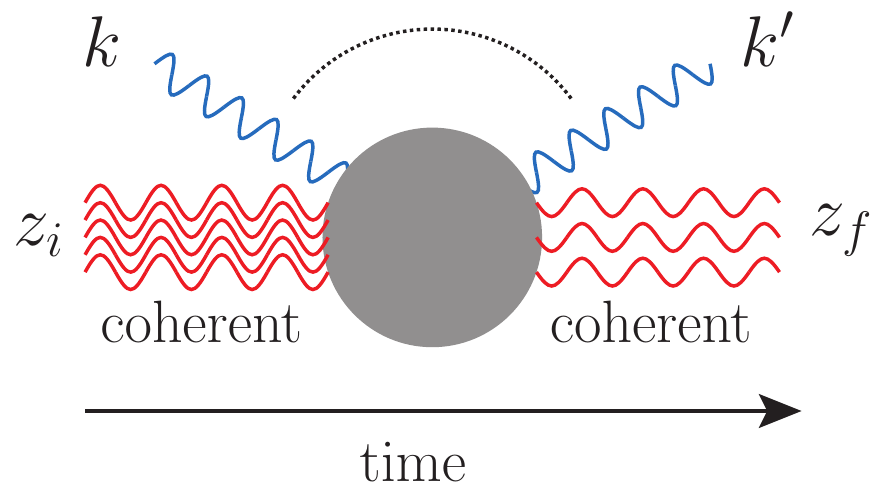}
\caption{\label{FIG:SPRIDNING} The physical situation of interest; an initial state containing matter (not shown), photons (in a number state) and a coherent state of photons (representing a strong field) scatters into a different state also comprising matter, emitted photons, and a coherent state of typically reduced amplitude representing depletion of the initial strong field.}
\end{figure}

We saw above that we can associate a background with asymptotic coherent states which are unaffected by scattering processes, i.e.~both the in and out states have the same coherent piece. This gives us a natural manner in which to include back-reaction on the background field; we allow the coherent state to change under the scattering process. The physical situation is illustrated in Fig.~\ref{FIG:SPRIDNING}. Our initial state contains incoming electrons and positrons (not shown), incoming photons (in a number state), and a coherent state of photons representing a strong field. In a real scattering situation we would typically take the initial coherent and number states to occupy well separated volumes of momentum space, i.e.~have different injection angles (in order to collide) and be of different frequencies (if e.g.~colliding high energy photons with an optical laser). Hence if $z_i({\bf k})$ is the profile function of the coherent state and $p$ the momentum of a photon in the number-state part, then we assume that $z_i({\bf p})=0$,
i.e. $[D(z_i),a^\dagger(\mathbf p)]=0$.
This assumption also means that there is no ambiguity in the definition of the initial state, for then
\be
	D(z) a^\dagger (\mathbf{p_1}) \ldots \ket{0} 
	= a^\dagger(\mathbf{p_1})\ldots D(z) \ket{0} 
	\equiv a^\dagger(\mathbf{p_1})\ldots \ket{z} \;,
\ee
which is normalised provided the number-state part is normalised.

Consider now the final state of the scattering process. During the interaction photons are emitted from particles, or absorbed from the initial coherent state. We describe scattered photons as a number state, as usual, and encode absorption from the initial state as a change in the coherent state profile, including a coherent piece $z_f({\bf k}) \not= z_i({\bf k})$ in the final state. We again assume that the momentum support of the final coherent state profile $z_f({\bf k})$ is disjoint from that of the \textit{final} number state. We do not make any assumption about the relative support of the initial coherent state and the final number state, or vice versa. Further, although we will not need to make this assumption explicitly, it is natural to imagine that $|z_f|^2 < |z_i|^2$ for some or all modes ${\bf k}$, representing losses from the initial field. In addition, the phase of the coherent state profile can be affected by the scattering process; identifying $z(\mathbf {k})$ with the Fourier components of the external field $A_\mathrm{ext}$, the phase $\varphi_\mathbf{k} := \arg z(\mathbf{k})$ resembles the spectral phase of the laser pulse. A change of this phase implies dispersion of the laser and could, for instance, strongly affect the pulse shape if certain frequencies were significantly `delayed'. This suggests an analogy with electrodynamics in continuous media, in which the number states act as the `medium' causing absorption and phase-shifts of the incident light.

We now turn to the task at hand, which is to calculate the scattering amplitudes describing the above. For clarity we will work mostly with scalar fields; this allows us to present a derivation uncluttered by spin, polarisation or gauge degrees of freedom. The extension of our final result to QED will though be straightforward. Hence let $\phi$ be a complex scalar representing either the scalar in sQED or the spinor in QED proper, coupled to a real scalar $A$ which represents the photon. The action is
\be
	S = \int\!\ud^4x\ \partial\phi^\dagger .\partial \phi - m^2\phi^\dagger \phi + \frac{1}{2}\partial A. \partial A  - \frac{1}{2}\mu^2 A^2+ S_\text{int}[A,\phi,\phi^\dagger] \;,
\ee
in which we include, for generality, a `photon' mass $\mu$. Although the interaction terms are in principle arbitrary (as is the dimension of spacetime, but we do not write this explicitly), for concreteness we will typically use the scalar Yukawa interaction
\be\label{ScalarYukawa}
	S_\text{int}[A,\phi,\phi^\dagger] = - e \int\!\ud^4 x \ A \, \phi^\dagger \phi \;,
\ee
in order to mock up the three-point vertex of QED, see also~\cite{Dybalski:2017mip}. (Note that $e$ then has mass dimension one in 3+1 dimensions, and is dimensionless in 5+1 dimensions.) The shortcomings of the scalar Yukawa theory will not concern us, as our initial interest is in establishing the formalism rather than extracting phenomenological results.

We require reduction formulae for scattering of particles between different coherent states of photons.  The mode expansion of the photon field reads
\be
	A(x) = \int\!\frac{\ud^3 {\bf k}}{\sqrt{(2\pi)^3 2\omega_{\bf k}}} \ a^\dagger({\bf k}) e^{ik.x} + a({\bf k}) e^{-ik.x} \;, \qquad [a({\bf p}),a^\dagger({\bf k})] = \delta^3({\bf p}-{\bf k}) \;.
\ee
For clarity we write $\mathcal{S}[A] \equiv \mathcal{S}[a,a^\dagger]$, making the photon creation and annihilation operators explicit but suppressing momentum labels. Then our goal is to calculate transitions of the form
\be\label{SFI-DEF-1}
	S_{fi}:=\bra{\text{out};z_f} \mathcal{S}[a,a^\dagger] \ket{\text{in};z_i} \;.
\ee
First we extract the coherent pieces from the initial and final states. We write out the coherent part of the initial state explicitly, and commute the exponential operator to the left using the translation property~(\ref{DaD}):
\be
	S_{fi} = \bra{\text{out};z_f} \mathcal{S}[a,a^\dagger] e^{a^\dagger z_i}\ket{\text{in}} e^{-|z_i|^2/2} =\bra{\text{out};z_f}  e^{a^\dagger z_i} \mathcal{S}[a+z_i,a^\dagger]\ket{\text{in}} e^{-|z_i|^2/2} \;.
\ee
(For clarity we write out only a single mode, the extension to all modes is trivial.) Now do the same for the final coherent state, passing it to the right of the $S$-matrix using the translation property and the standard results for the commutator of two displacement operators,
\be
\begin{split}\label{derive}
	S_{fi} &=  e^{-|z_f|^2/2} \bra{\text{out}} e^{{\bar z}_f a} e^{a^\dagger z_i} \mathcal{S}[a+z_i,a^\dagger]\ket{\text{in}} e^{-|z_i|^2/2} \\
	&= e^{-\frac{1}{2}(|z_f|^2 + |z_i|^2 - 2{\bar z}_f z_i) } \bra{\text{out}} e^{a^\dagger z_i} e^{{\bar z}_f a}  \mathcal{S}[a+z_i,a^\dagger]\ket{\text{in}} \\
	&= e^{-\frac{1}{2}(|z_f|^2 + |z_i|^2 - 2{\bar z}_f z_i) } \bra{\text{out}} e^{a^\dagger z_i} \mathcal{S}[a+z_i,a^\dagger + {\bar z}_f] e^{{\bar z}_f a}   \ket{\text{in}} \;.
\end{split}
\ee
The effect of the exponential operators on the initial and final states is to shift $a^\dagger({\bf p}) \to a^\dagger({\bf p}) + {\bar z}_f({\bf p})$ for each $a^\dagger$ in the in-state, and $a({\bf p}) \to a({\bf p}) + z_i({\bf p})$ for each $a$ in the out-state. We write these `shifted' states schematically as ``$\ket{\text{in}+{\bar z}_f}$'' and ``$\ket{\text{out}+z_i}$''. Finally, because $\mathcal{S}[a,a^\dagger]\sim \mathcal{S}[a e^{-ik.x} + a^{\dagger} e^{ik.x}]$, we see that the photon field in the $S$-matrix is once again shifted, but this time by a complex-valued background field $A_D$ for which $\bar{z}_f$ and $z_i$ are the negative and positive frequency modes respectively. Our key result is then:
\be\label{13}
	S_{fi} = e^{-\frac{1}{2}(|z_f|^2 + |z_i|^2 - 2{\bar z}_f z_i) } \bra{\text{out}+z_i} \mathcal{S}[A+A_D] \ket{\text{in}+{\bar z}_f} \;,
\ee
so that scattering between different incoming and final asymptotic states is equivalent to scattering in a complex-valued background, up to shifts in the initial and final states, and the leading factor in (\ref{13}) which can be recognised as the overlap of the asymptotic coherent states. While (\ref{13}) follows from basic properties of quantum mechanics, it gives us the main result on which the remainder of this paper is based.

That asymptotic coherent states lead to complex backgrounds has also been observed in~\cite{Zwanziger:1973if}. Indeed the appearance of a complex background should not be too surprising since complex potentials are commonly used as phenomenological models describing decay or absorption: complex optical potentials are an established way to describe inelastic scattering, as well as the energies and widths of resonances, see for example~\cite{Drisko:PhysLett1963,Georges:ChemPhys1986,Grazyna:PRA1983}, while imaginary potentials are used as a computational tool in numerical time-dependent Schr\"odinger equation (TDSE) simulations in order to absorb outgoing wavepackets at the boundaries of the simulation box~\cite{Muga:PhysRept2004}.

We stress that because we began with (\ref{SFI-DEF-1}), which is a transition amplitude of the ordinary, unitary, $S$-matrix between normalised states, unitarity is preserved despite the appearance of a complex background. As a check, we set $z_f = z_i$; we then recover the background field relation (\ref{vanligt}). Note that we need to assume no scattering into the beam, but this is consistent with the assumptions behind (\ref{13}); if $z_i(p_\text{in}) = z_f(p_\text{out})=0$ then because $z_i=z_f$ there can be no overlap between any of the number states and coherent states.

\subsection{Feynman rules}
The Feynman rules corresponding to the result (\ref{13}) are as follows.
\begin{enumerate}
	\item Correlation functions are derived from the shifted action
	\be\label{FEYN1}
	S = \int\!\ud^4x\ \partial\phi^\dagger .\partial \phi - m^2\phi^\dagger \phi + \frac{1}{2}\partial A. \partial A - \frac{1}{2}\mu^2 A^2 + S_\text{int}[A+A_D,\phi,\phi^\dagger] \;,
\ee
in which the photon field in the interaction term is shifted by a complex valued field $A_D$ which encodes both the incoming ($z_i$) and outgoing ($z_f$) coherent state profiles via
\be
		A_D = \int\!\frac{\ud^3 {\bf k}}{\sqrt{(2\pi)^32\omega_{\bf k}}} \ {\bar z}_f({\bf k}) e^{ik.x} + z_i({\bf k}) e^{-ik.x} \;.
\ee
(There is no shift in the free part of the action because the displacement operators in (\ref{derive}) act only on the interacting part of the Hamiltonian. See also Lecture 4 in~\cite{ZJ}.)
	\item Define the free asymptotic states (external photon legs) by
	\be\label{asympt-var}
		\varepsilon_p(x) := \frac{e^{-ip.x}}{\sqrt{(2\pi)^32 \omega_p}} \;, \qquad \omega_p := \sqrt{{\bf p}^2+\mu^2} \;.
	\ee
	For $S$-matrix elements insert the following into the path integral: for an incoming photon of momentum $p$, 
	\be\label{in-red}
{\bar z}_f({\bf p})+i\int\!\ud^4x\ \varepsilon_{p}(x)(\partial^2_x+\mu^2)\, A(x) \;,
\ee
and for an outgoing photon of momentum $q$, 
\be\label{ut-red}
	z_i({\bf q})+i\int\!\ud^4y\ \bar{\varepsilon}_{q}(y)(\partial^2_y+\mu^2)\, A(y) \;.
\ee
The second terms in these expressions describe the standard amputation of external legs with the Feynman $i\epsilon$ prescription, which replaces external propagators with the asymptotic wavefunctions (\ref{asympt-var}) and their conjugates. The first terms in (\ref{in-red}) and (\ref{ut-red}) correspond to the insertions in (\ref{13}) in which e.g.~a final state photon is sourced from the initial coherent state. Practically this means that for each insertion one has to calculate one additional Feynman diagram with one fewer photon leg, and multiply it by the appropriate coherent state profile function. While we will see that these terms may often be neglected, we will show in Sect.~\ref{SECT:MAXIMISERING} that they can play a crucial role in some circumstances. Similar terms occur in the infra-red safe LSZ reduction of charged particles and their radiation fields in~\cite{Zwanziger:1973if}. 
\item Multiply all {\it mod-squared} $S$-matrix elements by the real exponential
	\be\label{prefaktorn}
		\exp \bigg[- \int\! \ud^3{\bf k}\ |z_f({\bf k}) - z_i({\bf k})|^2\bigg] \,,
	\ee
as follows from (\ref{13}), and which accounts for the overlap and normalisation of the coherent states. (The imaginary part of the prefactor in (\ref{13}) drops out at the level of the probability.) When $z_f$ and $z_i$ are very different from each other it seems that this exponential will lead to a suppression of any probability or cross section; however, it should be remembered that the rest of the the amplitude depends in a highly nontrivial way on $z_f$ and $z_i$, so it is not possible to draw such general conclusions. We will look at examples below.
\end{enumerate}

This establishes our formalism and offers a potential framework in which to investigate back-reaction and depletion effects. Aside from the additional factors in LSZ reduction, the only essential difference between calculations in the above formalism and ordinary background field calculations is that the background has become complex valued.

\section{Basic examples}\label{SECT:EXEMPEL}
In this section we consider some simple illustrative examples of the above prescription. The emphasis is on establishing the applicability of the formalism, rather than on providing phenomenological results.
\subsection{Strong depletion of a weak field}\label{SECT:STARKSVAG}
Consider pair production, momenta $p$ and $p'$, by a photon of momentum $k'$. This process is forbidden by energy-momentum conservation in vacuum, but not if there is a background field present, see e.g.~\cite{Nikishov:1963,Heinzl:2010vg,Nousch:PLB2012}. We will compare the background field amplitude, $z_f=z_i$, or no depletion, with the case that the the field is completely depleted during the scattering (the initial coherent state is completely absorbed) so $z_f=0$.  We assume first that the field is weak, so that an entirely perturbative treatment is reasonable. Hence, rather than adopt the Furry expansion, we work for the moment in ordinary perturbation theory, expanding in powers of $e$ such that the coupling of the matter field to the complex field $A_D$ is treated on the same footing as the coupling to the quantised photon field.

Consider then the background field calculation with $z_f=z_i$. The background field is
\be\label{Aext}
	A_\text{ext}(x)  = \int\! \ud^3 {\bf k}\  \bar\varepsilon_k(x) \bar z_i({\bf k}) + \varepsilon_k(x) z_i({\bf k}) \quad \in\mathbb{R}\;,
\ee
with $\varepsilon_k(x)$ as in (\ref{asympt-var}). The lowest order contribution to the pair production amplitude $\mathcal{A}$ in this background is of order $e^2$, and is given by a superposition of standard ``two photon to pair'' or ``one photon to pair plus photon'' tree level Feynman diagrams $\mathcal{A}_0$ in which the one photon leg with momentum $k$ is attached to the background field $A_\text{ext}$:
\be
	\includegraphics[width=0.7\textwidth]{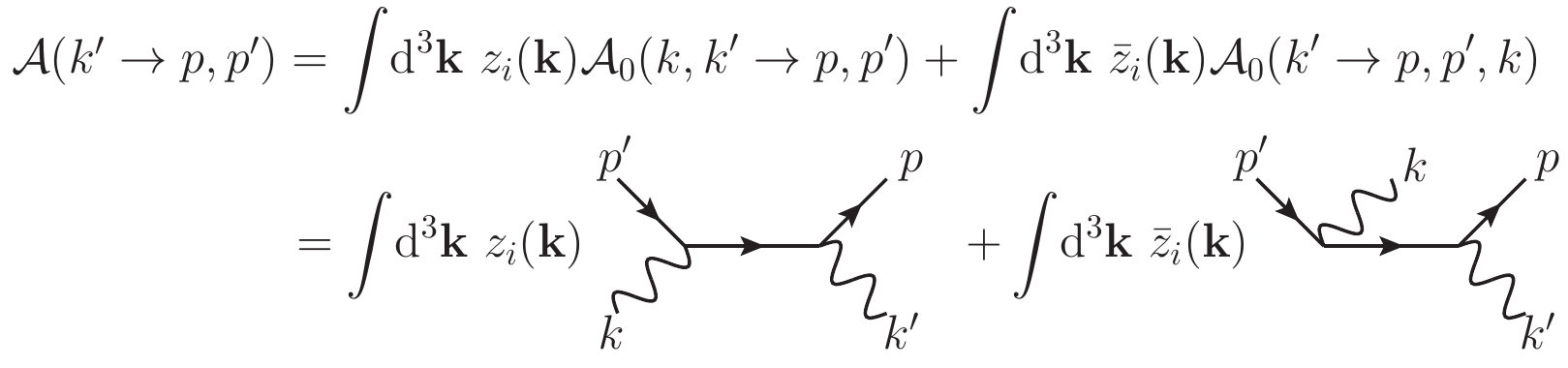} \ \raisebox{25pt}{.} 
\ee
The second diagram above vanishes by momentum conservation. Hence to this order we may freely replace the background field (\ref{Aext}) by 
\be\label{tillAD}
	A_\text{ext}(x)   \to  A_D(x) = \int\! \ud^3 {\bf k}\  \varepsilon_k(x) z_i({\bf k}) \;,
\ee
which is equal to the $A_D$ we would use in the case of complete depletion, $z_f=0$. It follows that, to lowest order in perturbation theory, the only difference between the amplitudes for no-depletion and complete depletion is the inclusion of the prefactor (\ref{prefaktorn}).  Then we have the simple result
\be
	\underset{\text{complete depl.}}{\mathbb{P}(k'\to p,p')} \simeq \exp\bigg[-\int\!\ud^3 {\bf k}\ |z_i({\bf k})|^2\bigg] \underset{\text{background}}{\mathbb{P}(k'\to p,p')} \;.
\ee
This result says that as the photon density $\sim |z|^2$ in the initial state increases, the probability of completely depleting the field during a perturbative process falls. This makes intuitive sense; we would not expect a field containing many photons to be entirely depleted by a single pair production event.

\subsection{Diagrammatic expansion}\label{DEXP}
\begin{figure}[t!]
	\includegraphics[width=0.7\textwidth]{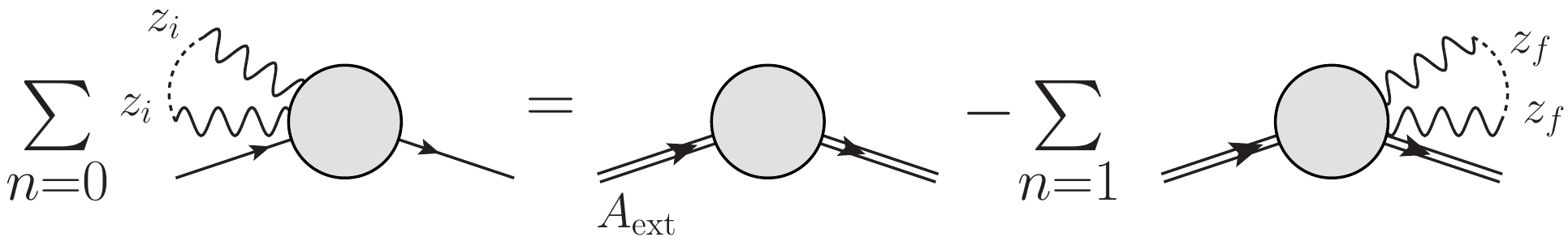}
	\caption{\label{FIG:AEXT-AD} Illustrative relation between two descriptions of the same process, in which an initial coherent state $z_i$ is completely depleted. On the left, the coherent state is expanded as a sum of incoming photons, none of which survive to the final state. On the right, the same expression is rewritten in terms of an infinite series of corrections to the background field amplitude, in which the corrections remove background field diagrams with photons in the final state.}
\end{figure}

In later examples we will go beyond perturbation theory and treat at least part of the field $A_D$ exactly. In preparation for this we now look at the diagrammatic representation of depletion effects. Consider again a field which is fully depleted, so that our complex $A_D$ is given by (\ref{tillAD}). Let $(A,\phi^\dagger\phi)$ be a condensed notation for the integral in the scalar Yukawa interaction~(\ref{ScalarYukawa}). Then the path integral describing some chosen scattering amplitude depends on $A_D$ through the term
\be\label{INT-TERMEN}
	e^{-i e (A_D,\phi^\dagger\phi)} = \sum\limits_{n=0}^\infty \frac{(-ie)^n}{n!} (A_D,\phi^\dagger\phi)^n \;,
\ee
which, given that $A_D$ in (\ref{tillAD}) has only positive frequency modes, describes a sum over all numbers of \textit{incoming} photons (connecting to $\phi^\dagger\phi$ at vertices in the scattering amplitude). This is natural, as we have no photons in the outgoing coherent state. Now, we can rewrite $A_D$ as the undepleted field (\ref{Aext}) together with a correction,
\be\label{delta-A-def}
	A_D = A_\text{ext} - \delta A \;,
\ee
where $\delta A$ has the form, comparing (\ref{Aext}) and (\ref{tillAD}),  
\be\label{dA}
	\delta A =  \int\! \ud^3 {\bf k}\  \bar\varepsilon_k(x) \bar z_i({\bf k}) \;,
\ee
and which describes only outgoing photons. With this (\ref{INT-TERMEN}) may be written as
\be\label{INT-TERMEN-2}
	e^{-i e (A_D,\phi^\dagger\phi)} = e^{-i e (A_\text{ext},\phi^\dagger\phi)} e^{i e (\delta A,\phi^\dagger\phi)} =  e^{-i e (A_\text{ext},\phi^\dagger\phi)} \sum\limits_{n=0}^\infty \frac{(-ie)^n}{n!} (-\delta A,\phi^\dagger\phi)^n \;.
\ee
This expression shows us how depletion effects look in the Furry expansion of a scattering process;          $A_\text{ext}$ remains in the exponent and is, implicitly, to be treated exactly using the Furry expansion.  Because this expansion is equal to (\ref{INT-TERMEN}), the series in $\delta A$ must therefore describe the depletion of $A_\text{ext}$: these terms \textit{subtract} from the background field process diagrams in which (all numbers of) photons are emitted into the final state (since $\delta A$ has only negative frequency modes, see (\ref{dA})). These diagrams cancel contributions which are included by $A_\text{ext}$, which contains both positive and negative frequency modes. It is for this reason that we have included a negative sign in the definition of $\delta A$; it emphasises that something is being removed. This is illustrated in Fig.~\ref{FIG:AEXT-AD}.

The series expansion \eqref{INT-TERMEN-2} will look very similar if we consider the more general case of of $z_f\neq0$, with the only
difference being that in this case we need to make the replacement, in~\eqref{dA} and \eqref{INT-TERMEN-2}, $\bar z_i \to \bar z_i - \bar z_f$ in $\delta A$. We will elaborate on this in the following section.

\subsection{Weak depletion of a strong field}\label{weak-strong}
Having understood the appearance of depletion effects, we now turn to the opposite limit as compared to that in Sect.~\ref{SECT:STARKSVAG}; we image that we have a strong field, and that this field is only mildly depleted during some given process. We take as an ansatz for the depleted field
\be
	z_f({\bf k}) = \big(1-\delta({\bf k})\big)z_i({\bf k}) \;,
\ee
in which $\delta({\bf k})\gtrsim 0$ is to describe a small reduction in amplitude of the final coherent state relative to the initial. We work to first order in $\delta$. In the path integral for the scattering amplitude we then expand the action as in (\ref{INT-TERMEN-2}),
\be\label{eps-expansion}
	\exp(i S) \to \exp(i S_\text{ext}) \bigg(1+i e\int\!\ud^4x\ \phi^\dagger\phi \int\! \ud^3{\bf k}\  {\bar \varepsilon}_k(x) {{\delta}({\bf k}){\bar z}_i({\bf k}) \bigg) \;.}
\ee
Here $S_\text{ext}$ is the background field action in which the matter fields are, as well as being coupled to the dynamical photon field, coupled to the background (\ref{Aext}). If the background is strong, this coupling should be accounted for in the Furry picture, as in Fig.~\ref{FIG:FURRY}. The effect of the $\mathcal{O}(\delta)$ term in (\ref{eps-expansion}) on the path integral is to insert into any process an additional (consistently normalised) outgoing photon line of momentum ${\bf k}$, and then to integrate this against {${\delta}({\bf k}){\bar z}_i({\bf k})$} which hence appears as a wavepacket.

So, let $\mathcal{S}_i^f$ be the Furry picture $S$-matrix element for some process $i�\to f$ including the scattering of photons with momenta $l_j$ to momenta $l'_j$, all in the {\it background} field (\ref{Aext}). The insertions of $z_i(l')$ and ${\bar z}_f(l)$ in (\ref{in-red}) and (\ref{ut-red}) may be dropped because e.g.~$z_f(l) \propto z_i(l) = 0$ for all $l_\mu$ by assumption. Then, expanding the general result in powers of~$\delta$, we find
\be
	\mathbb{P} = e^{-\delta^2\int |z_i|^2} \int\!\ud p_f\ \big| \mathcal{S}_i^f - \int\!\ud^3{\bf k}\ {\delta({\bf k}) {\bar z}_i({\bf k})} \, \mathcal{S}_i^{f,k} + c.c. + \mathcal{O}(\delta^2)\big|^2 \;,
\ee
where ${\mathcal S}_i^{f,k}$ is the $S$-matrix element for the process same process as in ${\mathcal S}_i^{f}$ but with the emission of an additional photon into the final state. As described above and illustrated in Fig.~\ref{FIG:AEXT-AD}, this diagram describes the depletion of the background field $A_\text{ext}$. To lowest order in $\delta$, the exponential pre-factor does not contribute, and the probability becomes
\be
	\mathbb{P} = \int\!\ud p_f  \big| {\mathcal S}_i^f \big|^2 - \int\!\ud p_f  \int\!\ud^3{\bf k}\  \bigg({ \delta ({\bf k})\overline{\bar z}_i({\bf k})}\, \mathcal{S}_i^{f,k}\, \overline{\mathcal{S}_i^f} + c.c. \bigg) + \mathcal{O}(\delta^2)\;.
\ee
We thus see that a reduction in amplitude of the coherent state corresponds, at the level of the probability or cross section, to cross-terms of different photon-number processes; one is the background field amplitude, the other is the amplitude for the same process but with an additional emission, integrated over the profile of the background field. This is in agreement with the discussion in Sect.~\ref{DEXP} and with Fig.~(\ref{FIG:AEXT-AD}).  To be explicit, consider again pair production by a photon of momentum $k'$. Writing a double line for the matter propagator in the background (\ref{Aext}) as in Fig.~\ref{FIG:FURRY}, and adopting the scalar-Yukawa interaction (\ref{ScalarYukawa}) we have
\be
	\includegraphics[width=0.5\textwidth]{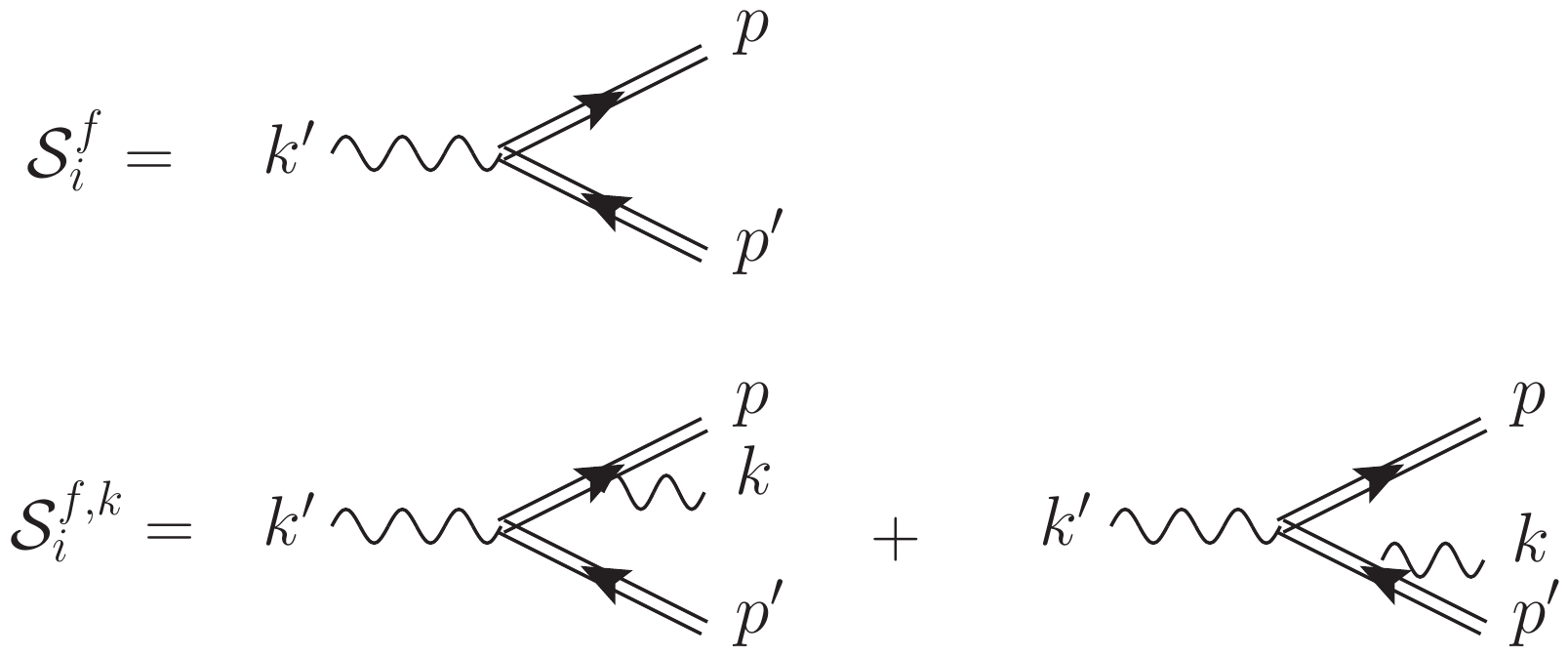}
\ee
We comment that the inclusion of additional diagrams giving higher-order photon emission is reminiscent of the IR, though the interpretation here is different. In our case the additional diagrams subtract photon emission contributions from the background field amplitude in order to describe a depletion of that field.

\subsection{Strong depletion of a strong field}\label{strongstrong}
\begin{figure}[t!]
\includegraphics[width=0.6\textwidth]{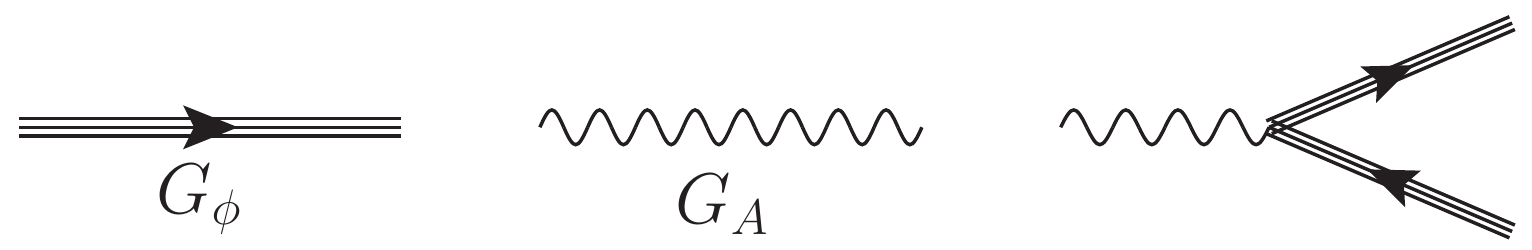}
\caption{\label{FIG:FEYN} Position-space Feynman rules in the Furry picture; the interaction between the matter fields and the coherent states (described by $A_D$) is accounted for exactly by using the propagator $G_\phi$ in (\ref{G-DEF}).}
\end{figure}

Consider the action (\ref{FEYN1}) together with, having QED in mind, the Scalar Yukawa interaction (\ref{ScalarYukawa}). Following the background-field calculations in the literature, see Sect.~\ref{SECT:review}, it is tempting to use a Furry picture expansion of a general correlation function in which the matter field propagator $G_\phi$ obeys
\be\label{G-DEF}
	G_\phi^{-1} = \epsilon +i(\partial^2+m^2+ e A_D) \;.
\ee
If $G_\phi$ can be found explicitly then the coupling to both the initial and final coherent states can be treated without approximation. In this approach particles propagate in the complex background field $A_D$ while the interaction (the three-point vertex) between the matter field and the quantised photon field is treated in perturbation theory as usual.  The Feynman rules are illustrated in Fig.~\ref{FIG:FEYN} and are precisely as for the usual Furry picture in Fig.~\ref{FIG:FURRY} except that the background field in the matter propagator is complex. Applying LSZ amputation (analogous to (\ref{in-red}) and (\ref{ut-red}) but without the additional terms) to external $G_\phi$ lines gives the asymptotic particle wavefunctions. 

There are some conceptual issues to be addressed here, which are best illustrated by example. Assume that both the initial and final coherent states are plane waves depending on $n.x$ for $n^2=0$. We should then in principle be able to make as much analytic progress in scattering calculations as in the background field case, as $G_\phi$ and hence the external leg wavefunctions can be written down exactly; the propagator (\ref{G-DEF}) for a plane wave background is
\be\label{G-AD}
	G_\phi(x,y) = \int\!\frac{\ud^4 p}{(2\pi)^4} \frac{1}{\epsilon - i (p^2-m^2)} \exp\bigg[-ip.(x-y) - \frac{i}{2n.p}\int\limits_{n.y}^{n.x} e A_D\bigg] \;,
\ee
which we recognise as being very similar to that in (scalar) QED, see e.g.~\cite{DiPiazza:2011tq,Ilderton:2012qe,Seipt:2017ckc} and references therein. Applying the standard analogues of (\ref{in-red}) and (\ref{ut-red}) to an external leg of a correlation function transforms the propagator into the following ``asymptotic wavefunctions'' analogous to the Volkov solutions in QED, albeit with a complex classical field: 
\be\begin{split}\label{volkov-ish}
	 \sqrt{(2\pi)^3 2 E_{\bf p}} \, e^\LCm_\text{out}(x) &=\int\!\ud^4 y\, e^{ip.y}(\epsilon+i(\partial^2_y+m^2)) G_\phi(y,x) =  \exp\bigg[ ip.x-\frac{i}{2n.p}\int\limits^{\infty}_{n.x} e A_D\bigg] \;, \\
	 \sqrt{(2\pi)^3 2 E_{\bf p}} \, e^\LCm_\text{in}(x) &=\int\!\ud^4 y\, e^{-ip.y}(\epsilon+i(\partial^2_y+m^2)) G_\phi(x,y) = \exp\bigg[ -ip.x-\frac{i}{2n.p}\int\limits_{-\infty}^{n.x} e A_D\bigg] \;,
\end{split}
\ee
and similarly for $e^\LCp$. These functions obey the Klein-Gordon equation in the background $A_D$. Their interpretation as one-particle wavefunctions is though unclear. The physical meaning of a particle propagating in a complex potential is not obvious, and the wavefunctions are not normalisable due to the imaginary part of $A_D$. Note also that $\bar A_D$ never occurs, because the $S$-matrix depends only on $A_D$.

It seems clear though that a perturbative solution of (\ref{G-DEF}), resummed to all orders in the coupling, would yield precisely (\ref{G-AD}) and from there (\ref{volkov-ish}), just as it does for real plane waves. Thus we could attempt to perform scattering calculations in this Furry-like picture.  Due to the assumption of plane wave coherent profiles there would be some divergent factors to assign a meaning to in e.g.~(\ref{prefaktorn}) corresponding to longitudinal and transverse volumes and numerical integration might be made harder by the presence of real exponentials rather than just phases. For these reasons we defer a full calculation in QED proper to future research, though we will briefly look at this plane wave model again below.

\section{Extremisation}\label{SECT:MAXIMISERING}
In Sect.~\ref{SECT:LSZ} we introduced the final state profile $z_f$ as a new degree of freedom describing the depletion of the laser. In Sect.~\ref{SECT:EXEMPEL} we fixed $z_f$ in order to model a chosen level of depletion, and explored how scattering probabilities depended on this. Now, in background field calculations there are two couplings, one between the quantised fields and one to the background field $\sim e A_\text{ext}/m$. The inclusion of depletion effects introduces (at least) a third parameter, the amount of depletion, which may be re-interpreted as the coupling to $A_D-A_\text{ext}$. In fixing $z_f$ as we did above, we made both explicit and implicit assumptions on the relative strengths of the three couplings. 

Here we take a different view, and instead try to determine the (most likely) level of depletion in a given process using an extremisation principle, much like the Hamiltonian action principle in classical mechanics (where the classical path is obtained by extremising the action) or the principle of extremisation of entropy. While we do not aim to imply that there is a strict analogy, extremisation principles are extremely useful in many different areas of physics~\cite{book:Sieniutycz}.

By fixing the initial coherent $z_i$ and the scattered particle content (as above), and then maximising the probability of a process with respect to the hitherto unknown $z_f$, we eliminate the dependency on the third coupling. The determined value of $z_f$ then represents the physically relevant `most likely' level of depletion. The goal is to be able to use this, eventually, to learn how depletion in QED scattering processes affects the spectrum of final state particles and of the laser pulse itself, giving possibly measurable signatures. In this section we investigate the feasibility of such an approach. We begin with a simple, but exactly solvable, model.

\subsection{Exactly soluble toy model}
\begin{figure}[t!]
\includegraphics[width=8cm]{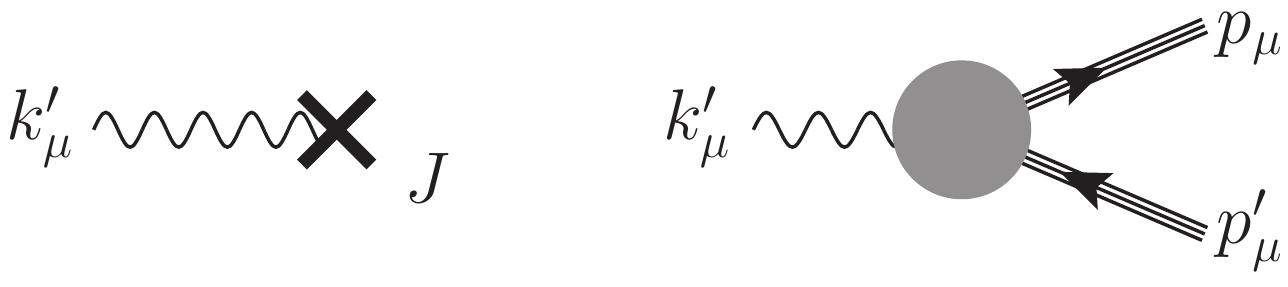}
\caption{\label{FIG:PARBILDNING} On the left, a photon is absorbed by the external source $J$.  This source mocks up the matter current on the right, in which a (scalar) photon produces a (scalar) pair in the Yukawa theory. The triple line indicates that the coherent states can in principle be accounted for exactly in the Furry picture and the blob represents loop corrections.}
\end{figure}

Consider a real scalar field $A$ (the `photon'  from the scalar Yukawa model) coupled to a given external source/sink~$J$. Including the complex field describing the asymptotic coherent states of the $A$-field, the action is 
\be
	S = \frac{1}{2}\int\!\ud^4 x \ \partial A .\partial A + \int\!\ud^4x \ J \big( A + A_D \big) \;,
\ee
with $A_D$ as in (\ref{Aext}). Since the scalar field couples linearly to the source the presence of $A_D$ in the action only trivially affects scattering processes, given a further exponential prefactor. There are however the insertions of (\ref{in-red}) and (\ref{ut-red}) to account for. Consider then the simple scattering process shown in the left hand diagram of Fig.~\ref{FIG:PARBILDNING}, in which an incoming photon of momentum $k'_\mu$ is absorbed by the source $J$. 
Let the initial state contain (as well as the coherent state $z_i$ accounted for in $A_D$) a single photon in a wavepacket $\psi$,
\be
	\ket{\psi} = \int\!\ud^3{\bf k}\ \psi({\bf k}) a^\dagger({\bf k}) \ket{0} 
	\;, \qquad \bracket{\psi}{\psi} = \int\! \ud^3{\bf k}\, |\psi({\bf k})|^2 = 1 \;.
\ee
We assume as before that the wavepacket $\psi$ is disjoint from $z_i$ in momentum space. The probability that the photon is absorbed by the source, together with a transition from the coherent state $z_i$ to the coherent state $z_f$ is now easily calculated (see e.g.~\cite{book:Itzykson} for details), remembering to include the additional terms in (\ref{in-red}). Defining the on-shell Fourier transform of $J$,
\be
	J({\bf k}) := \int\!\ud^4 x \  {\bar \varepsilon}_k(x)  \, J(x) \iff {\bar J} ({\bf k}) := \int\!\ud^4 x \  {\varepsilon}_k(x)  \, J(x) \;,
\ee
the total photon absorption probability is, without approximation,
\be\label{P-parbildning-enkel-1}
	\mathbb{P}(z_f) = e^{-\int \!\ud^3{\bf k} \, |z_f-z_i- i J|^2} \bigg| \int\! \ud^3{\bf k}\ \psi({\bf k})\big[\bar{z}_f({\bf k}) + i \bar{J}({\bf k}) \big] \bigg|^2 \;,
\ee
where the insertion of $\bar{z}_f$ arises from (\ref{in-red}). The coherent-state normalisations and the contribution of $A_D$ from the action (and from the conjugate $S$-matrix element) combine into a single exponential.  From this we can identify that final $z_f$ which maximises the probability. Extremising (\ref{P-parbildning-enkel-1}) with respect to $\bar{z}_f({\bf p})$ yields
\be\label{zf-eqn}
	\frac{\delta \mathbb{P}}{\delta {\bar z}_f({\bf k})} = 0 \implies z_f({\bf k}) = i J({\bf k})  \quad \text{or}\quad \big(z_f({\bf p}) - z_i({\bf p}) -i J({\bf p})\big) \int\!\ud^3{\bf k}\, \psi({\bf k})\big[\bar{z}_f({\bf k}) + i \bar{J}({\bf k}) \big] = \psi({\bf p}) \;.
\ee
The first condition minimises the probability, setting it to zero. The solution to the second condition is (remembering that there is no overlap between $z_i$ and $\psi$), 
\be\label{zf-soln}
	z_f({\bf k}) = z_i({\bf k}) + i {J}({\bf k}) + e^{i\theta} \psi(\bf{k}) \;,
\ee
in which $\theta$ is an arbitrary, but constant, phase implying, recall the earlier discussion, a possible spectral phase shift in the final coherent state profile. The $iJ$-term simply describes (as we will make explicit below) the standard instability of the empty vacuum state in the presence of the external source $J$~\cite{book:Itzykson}. The $\psi$ term comes from the LSZ insertion in (\ref{in-red}) and means, interestingly, that the support of the \text{final} coherent state profile overlaps with that of the \textit{initial} number state. If we associate $J$ with some coupling, then (\ref{zf-soln}) shows us that the final coherent state is corrected not only at first order in the coupling, due to the interaction, but also at zeroth order through the initial state $\psi$. Inserting the solution (\ref{zf-soln}) into (\ref{P-parbildning-enkel-1}) gives a maximal probability of
\be\label{P-parbildning-enkel-2}	
	\mathbb{P} \to \mathbb{P}_\theta =   e^{-\int\!\ud{\bf k} |\psi|^2} \bigg| \int\!\ud{\bf k}\, |\psi|^2 \bigg|^2  = e^{-1} \;,
\ee
independent of $\theta$. Despite the simplicity of the theory, this example illustrates that it is possible to maximise the probability with respect to the final coherent state profile. The implied choice of final state depends both on the interaction and on the properties of the initial state.

Note that we have not insisted that the final coherent state amplitude be reduced with respect to the initial; the final state which maximises the probability could contain, as well as absorption from the initial state, coherent emission into the final state.  Indeed, the assumption that the initial number and coherent state parts are disjoint in momentum space allows us to write
\be
	z_f({\bf k}) = \begin{cases}
				z_i({\bf k}) + i J({\bf k}) & \text{when } z_i({\bf k}) \not=0  \;, \\
				e^{i\theta}\psi({\bf k}) + i J({\bf k}) & \text{when } z_i({\bf k}) =0 \;,
				\end{cases}
\ee
so that along with the population of coherent state modes due to $J$, which is particular to this model, we see that populated (initial) number state modes become populated (final) coherent state modes.  It seems clear that there can be physical situations in which emission is more significant than absorption, and vice versa. The expectation value of the number of photons in the final state, $N_f$, is given by 
\be\begin{split}
	N_f = \int\!\ud^3{\bf k}\, |z_f({\bf k})|^2 &= 1+ N_i  + \int\!\ud^3{\bf k}\, |J({\bf k})|^2 +   \int\!\ud^3{\bf k}\, i  J({\bf k}) (\bar{z}_i({\bf k}) + e^{-i\theta}\bar\psi({\bf k})) - i \bar{J}({\bf k}) (z_i({\bf k}) + e^{i\theta}\psi({\bf k}))\;,
\end{split}
\ee
where, in contrast to the probability above, the phase survives. If we average over the phase then we can write
\be\begin{split}
	\int\limits_0^{2\pi}\! \frac{\ud\theta}{2\pi} N_f = 1+ \int\!\ud^3{\bf k}\, |z_i({\bf k}) + i J({\bf k})|^2 \;.
\end{split}
\ee
Whether the final state corresponds to an overall reduction or increase in the average number of photons therefore depends on the distance (in the complex plane) between the initial state and the source.

We comment that the expectation value of $z_f$ in the final state is
\be
	\langle z_f({\bf k}) \rangle = \int\!\pathD z_f \ z_f({\bf k}) \mathbb{P}(z_f) = z_i({\bf k}) + i J({\bf k}) \;,
\ee
from which it follows that the expectation value of $A(x)$ in the asymptotic future is (remembering to convert back from the interaction picture)
\be\begin{split}
	\langle A(x) \rangle &= A_i(x) + i \int\!\frac{\ud^3 {\bf k}}{(2\pi)^3 2\omega_k} \big(e^{-ik.(x-y)}- e^{ik.(x-y)}\big) J(y) = A_i(x) + \int\!\ud^4y\, G_\text{rad}(x-y) J(y) \;,
\end{split}
\ee
in which $G_\text{rad} = G_\text{ret}- G_\text{adv}$ gives the radiation field~\cite{Zwanziger:1973if}. This comes entirely from the $iJ$ term in (\ref{zf-soln}), i.e.~is independent of $\psi$, and is the same as the result obtained if starting from vacuum.  That this is independent of the initial number state properties is due to the simplicity of the model, as coherent states evolve directly to coherent states when the only interaction is with an external source. We would certainly expect, in general, that if large numbers of particles were present then interactions, and hence absorption, would be more likely. This is correctly reflected by~(\ref{zf-soln}).

\subsection{Pair production in scalar Yukawa theory}
We return to scalar Yukawa and consider again stimulated pair production by an initial photon. 
This is the natural generalisation of the absorption process in the toy model above, see Fig.~\ref{FIG:PARBILDNING}. Let the initial state be a single photon with wavepacket $\psi({\bf k}')$, while the final state contains a pair (of matter particles) with momenta $p$ and $p'$. Write $\text{Amp}_{{\bf k}'}[A]$ for the amputation instruction for photons, and define $\text{Amp}_{\bf p}[\phi]$ similarly for the matter fields but without additional insertions corresponding to $z_i$ or $z_f$ as the matter states contain no coherent pieces. Then the probability of stimulated pair production is
\be\label{pathint}
\begin{split}
	\mathbb{P} &=  \int\!\ud p_f  \: e^{- \int \! \ud k | z_f-z_i |^2 } \, \big|\mathcal{M}\big|^2 \;, \\
	\mathcal{M} &= \int\! \ud^3{\bf k'}\ \psi({\bf k'}) \int\!\pathD\phi\, \pathD \phi^\dagger\,\pathD A\,  \text{Amp}_{{\bf k}'}[A]  \text{Amp}_{{\bf p}}[\phi^\dagger] \text{Amp}_{{\bf p}'}[\phi] \, e^{i S[\phi,\phi^\dagger,A,A_D]} \;,
\end{split}
\ee
in which $\ud p_f = \ud^3{\bf p}'\ud^3{\bf p}$ is the final state integral over outgoing pair momenta.

Note that the exponential factor containing the difference of asymptotic coherent state profiles stands under the phase space integral.  The reason is the following. Scattering to a final state with disjoint number and coherent pieces has been shown in~\cite{Frantz} to be equivalent, at the level of the probability, to an \textit{inclusive} process in which one sums over all numbers of photons emitted into a `forward' direction given by the support of the coherent state. With this in mind, it makes sense for the level of depletion in each mode of $z({\bf k})$ to be allowed to depend on the momenta of the final state particles, and for this reason the exponential stands under the final state integrals.

Differentiatiating with respect to $\bar z_f({\bf k})$ we obtain
\begin{align}\label{pathint2}
\frac{\delta \mathbb P}{\delta \bar z_f(\mathbf{k})} 
		= \int\!\ud p_f  \: e^{- \int \! \ud k | z_f-z_i |^2 } \,
		\left(
				 \overline{\mathcal M}  \frac{\delta \mathcal M}{\delta \bar z_f(\mathbf{k})} 
		 	- 		 (z_f(\mathbf{k}) - z_i(\mathbf{k}))   |\mathcal M|^2 
 		\right) \,.
\end{align}
Demanding that that this expression be extremised for each momentum mode then requires us to solve the deceptively simple equation
\begin{align} 
\label{tosolve}
z_f(\mathbf{k}) = z_i(\mathbf{k})  + \frac{1}{\mathcal M} \frac{\delta \mathcal M}{\delta \bar z_f(\mathbf{k})} \,,
\end{align}
which generalises (\ref{zf-eqn}) and \eqref{zf-soln}. This time, however, \eqref{tosolve} is a complicated, highly non-linear equation for the final coherent state profile $z_f$, since the matrix element $\mathcal M$ depends on $z_f$ through the complex background $A_D$. Clearly we cannot solve this new equation exactly, so the simplicity of the extremised state distribution (\ref{zf-soln}) and probability (\ref{P-parbildning-enkel-2}) will not extend to this system. We will though still be able to write down a series solution to (\ref{tosolve}). The extension to QED will also be apparent.

We write $A_D = A_\text{ext} - \delta A$ as in (\ref{delta-A-def}), and treat the coupling to the background field $A_\text{ext}$ exactly while treating the coupling to the quantised $A$ field and to the fluctuation $\delta A$ in perturbation theory, see (\ref{INT-TERMEN-2}), i.e.~we work in the ordinary Furry picture. Before proceeding to solve (\ref{tosolve}) in this expansion, let us establish some notation. Define the two background field amplitudes
\be\label{EQN:SFI2}
	\includegraphics[width=9cm]{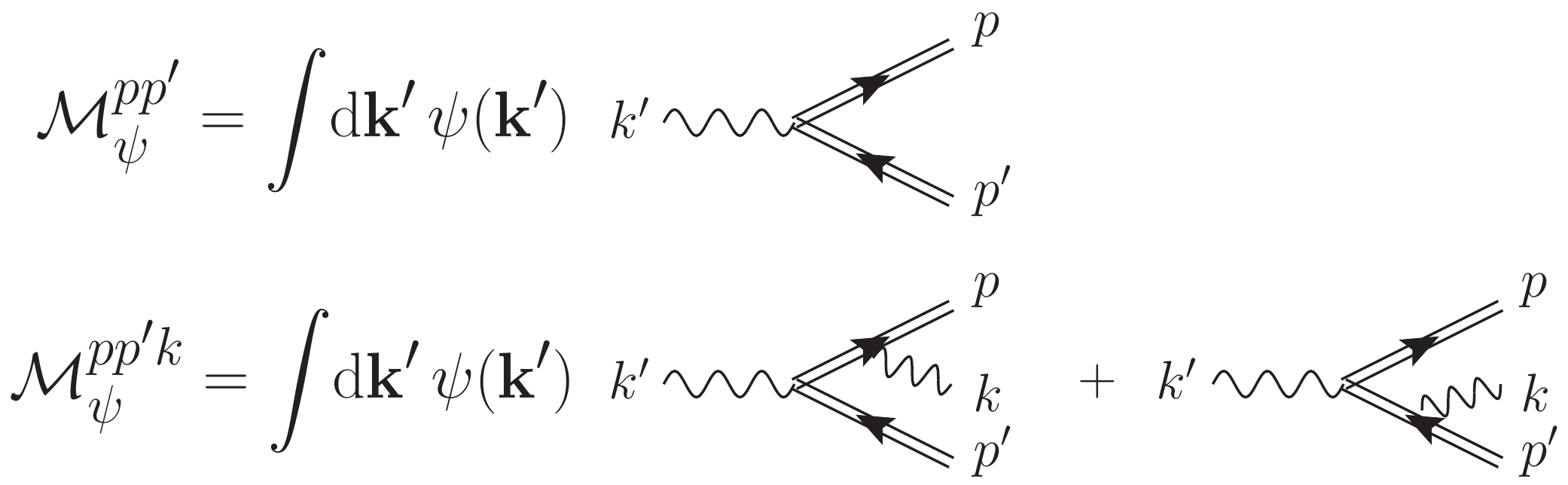} \;\; \raisebox{15pt}{,}
\ee
which differ from the diagrams considered earlier only in the inclusion of the initial wavepacket. It is clear why the first is relevant; the second arises here because the effect of the $z_f$-derivative in (\ref{tosolve}) is to insert into the path integral defining $\mathcal M$ a term
\be
	-ie \int\!\ud^4x \ \bar\phi \phi\ \bar\varepsilon_k(x) \;,
\ee
which just attaches a new, correctly normalised, outgoing photon line to our process; we saw such terms previously in (\ref{eps-expansion}).  Since depletion is here being caused by an order $e$ process we make the ansatz $z_f = z_i + \delta z_0 + e \delta z + \ldots$ and neglect higher powers in $e$. Here the zeroth order correction is suggested by the exactly solvable model above. Things simplify somewhat if we assume that the background field $A_\mathrm{ext}$ is incapable of spontaneous (i.e.~nonperturbative, or Schwinger) pair production. In that case the LSZ insertion $\bar z_f$ in $\text{Amp}[A]$ does not contribute, and further one finds that $\delta z_0 = 0$. Thus we have $z_f = z_i + e \delta z$, and one finds
\begin{align}
 e \delta z({\bf k}) =  \frac{\mathcal M^{pp'k}_\psi}{\mathcal M^{pp'}_\psi}\,,
\end{align}
where both the background field amplitudes from (\ref{EQN:SFI2}) appear. Thus we have demonstrated that the extremisation principle can be successfully applied to scalar Yukawa theory. One can proceed to higher orders. 

We remark briefly that one could also also look for an `overall' level of depletion which depends only on integrated emission rates, not on the differential rates. Here one returns to (\ref{pathint}) but places the exponential outside of the final-state integrations. Then extremising with respect to $\bar z_f({\bf k})$ yields
\be\label{tosolve2}
	(z_f({\bf k})-z_i({\bf k})) \int\!\ud p_f\ |\mathcal{M}|^2 = \int\!\ud p_f\ \frac{\delta \mathcal{M}}{\delta z_f({\bf k})} \overline{\mathcal{M}} \;.
\ee
Proceeding as above, the zeroth order correction vanishes while the first nontrivial correction, $e \delta z_1$ is
\be\label{delta-z-1-soln}
	e \delta z_1({\bf k}) = \frac{\displaystyle\int\!\ud p_f \overline{\mathcal{M}_{\psi}^{pp'}}\mathcal{M}_{\psi}^{pp'k}}{\displaystyle\int\!\ud p_f \big|\mathcal{M}_{\psi}^{pp'}\big|^2} \;.
\ee
Here we see the same cross terms arises as in earlier sections. There is a simple relation between these two approaches. We can consider the inclusive depletion of a mode $z(\mathbf{k})$ when the scattered particles are `not observed' by integrating $e\delta z$ over the final state phase space, weighted by the normalised (background field) differential scattering probability, $(1/\mathbb{P})(\ud \mathbb P/\ud p_f)$. Then we find 
\be
\int \! \ud p_f \: e \delta z({\bf k}) \: \frac{1}{\mathbb P}\frac{\ud \mathbb P}{\ud p_f} = e \delta z_1({\bf k}) \;. 
\ee 

\subsection{Depletion of a plane wave}\label{PWDEP}
Finally, let us take another step toward including depletion effects in the strong field QED calculations of the literature, by considering the case that the initial coherent state is a plane wave, with support only on momentum modes in a single (lightlike) direction $n_\mu$. We have seen that the first correction to the initial state profile involves the amplitude $\mathcal M^{pp'k}_\psi$, in which a photon is emitted along with the created pair. Overall momentum conservation for this amplitude takes the form
	\be
		k'_\mu  + \omega n_\mu = p_\mu + p'_\mu + k_\mu \;,
	\ee
	where $\omega$ is some frequency scale determined by the mass-shell condition. The point to make is there are solutions in which the emitted photons can have momenta $k_\mu \not\propto n_\mu$ and hence the final coherent state profile $z_f = z_i + e \delta z$ does not necessarily depend only on $n.x$, i.e.~is not necessarily a plane wave. This shows that including higher order depletion effects will ultimately also require going beyond plane waves models of the background, which is another challenging area of research~\cite{DiPiazza:2016maj,Heinzl:2017zsr}.

Out of interest, though, to see what kind of structures may arise in a more complete investigation, we impose the assumption of Sect.~\ref{strongstrong} that both the initial and final coherent states are plane waves, characterised by the lightlike direction $n_\mu$. This allows us to make a little more analytic progress. Because both the initial and final state profiles depend on the same lightlike direction only, the variation of $\mathcal{M}$ with regard~to ${\bar z}_f$ simplifies, and reduces to the variation of $\mathcal{M}_\psi^{pp'}$ itself. If we write $z({\bf k}) = \delta^2({\bf k}_\LCperp)\theta(k_3) \zeta(k_3)$ for each coherent state profile, then the equation to solve for the first correction to $\zeta$ becomes
	\be\label{tosolvezeta}
		e \delta \zeta(k) V_\LCperp = \frac{1}{\mathcal{M}_\psi^{pp'}} \frac{\delta \mathcal{M}_\psi^{pp'}}{\delta \bar\zeta(k)} \bigg|_{\zeta = \zeta_i} \;,
	\ee
	where the volume factor arises from differentiating the exponential in (\ref{pathint}). So, we first calculate the matrix element using the Volkov-like states~\eqref{volkov-ish} with a background field $A_\text{ext}$. As is usual in plane wave calculations, momentum is conserved in three directions which, in this case, allows us to perform the three integrals over the wavepacket momenta $k'$. Let us collect the resulting wavepacket factor together with all normalisation factors (e.g.~factors of $\sqrt{2\omega}$ and so on), into a single factor $\mathcal C$. The nontrivial part of $\mathcal{M}_\psi^{pp'}$ is an integral over the lightlike direction $n.x \equiv x^\LCp$ on which the external field depends,
\begin{align}
	\mathcal{M}_\psi^{pp'} & = \mathcal{C} \int \! \ud x^\LCp \: \exp\bigg[ i K_\LCp x^\LCp - i \bigg(\frac{e}{2n.p} + \frac{e}{2n.p'}\bigg) \int\limits_{x^\LCp}^{\infty}\! \ud s\, A_\text{ext}(s) \bigg] \\
	&\equiv \mathcal{C}\, J(K_\LCp)\;,
\label{eq:ppSY-J}
\end{align}
in which we have defined $K_\LCp = p_+ + p'_+ -k'_+$ (all evaluated on-shell), and $J$ is a `transition current' which may be thought of as a generalisation of the classical source ($J$) above. Taking the derivative in (\ref{tosolvezeta}), the final result becomes (observing that for our chosen plane wave $k_\LCp = k_3$ and so $k_\mu = k_\LCp n_\mu$),
\be\label{eq:zf-final}
	e \delta \zeta(k_\LCp) = \frac{1}{V_\LCperp} \bigg(\frac{e}{2k.p} + \frac{e}{2k.p'}\bigg) \frac{1}{\sqrt{(2\pi)^32k_\LCp}}\frac{J(K_\LCp + k_\LCp)}{J(K_\LCp)} \;.
\ee
We see explicitly from~\eqref{eq:zf-final} that the change in laser mode $k_+$ depends also on the momenta of the final number state particles, through $K_+$. The terms in round brackets follow from taking the functional derivative and correspond, recall (\ref{EQN:SFI2}), to the emission of the additional photon from either the final electron or positron line. Interestingly, these terms are strongly reminiscent of infra-red factors in QED. Although the interpretation may be different, it seems that the appearance of depletion effects in QED will have a lot in common with infra-red structures and higher-order processes. While the infra-red is often assumed to be well understood, there is still a great deal of interesting physics to explore there, see~\cite{Lavelle:1995ty,Horan:1999ba,Lavelle:2005bt,Kapec:2017tkm} and references therein. Further, the asymptotic behaviour of particles in plane waves and their associated IR divergences~\cite{Dinu:2012tj,Ilderton:2012qe} has a reinterpretation as an electromagnetic analogue of gravitational memory effects~\cite{Bieri:2013hqa,Adamo:2017nia}.

\section{Discussion and Conclusions}\label{SECT:CONS}
%
\subsection{Extension to QED}
The extension of our basic formalism to QED is straightforward. In Lorenz gauge $\partial_\mu A^\mu = 0$ the only changes needed are to keep track of signs coming from the metric and vector structure. Recall that the canonical commutation relation for the photon creation and annihilation operators is, note the sign,
\be
	\big[a_\mu({\bf p}) , a^\dagger_\nu({\bf q})\big] = - g_{\mu\nu} \delta^3({\bf p}-{\bf q}) \;.
\ee
It follows that the prefactor multiplying mod-squared $S$-matrix elements becomes
\be
	\exp \left[ + \int\!\ud^3 {\bf k}\ (z_f - z_i)_\mu (\bar{z}_f - \bar{z}_i)^\mu \right]\;.
\ee
(This prefactor can be written in terms of the difference of two potentials, 
$A^\mu_\text{out}(x) - A^\mu_\text{in}(x)$, and is therefore gauge invariant.) The QED action itself becomes
\be\begin{split}
	S &= \int\!\ud^4x \ \bar{\psi} \big(i\slashed{\mathcal D} - m \big) \psi - \frac{1}{4}F_{\mu\nu} F^{\mu\nu} - e \bar\psi \slashed{A} \psi + \text{gauge fixing} + \text{ counterterms,} \\
	{\mathcal D}^\mu &:= \partial_\mu + i e A^\mu_D(x) \;,
\end{split}\ee
in which $A^\mu_D$ is now the complex valued field
\be
	A^\mu_D = \int\!\frac{\ud^3 {\bf k}}{\sqrt{(2\pi)^3 2\omega_{\bf k}}} \ {\bar z}^\mu_f({\bf k}) e^{ik.x} + z_i^\mu({\bf k}) e^{-ik.x} \;.
\ee
As a first investigation in QED it would seem natural to follow the ideas of Sect.~\ref{weak-strong} and Sect.~\ref{PWDEP} in which we perturb around external field calculations assuming the plane wave model, as that is a well-established formalism. This will be investigated elsewhere.

\subsection{Conclusions}
Within the framework of `quantum field theory in background fields' one can identify (at least) two primary theoretical challenges. The first is to accurately account for realistic, typically complex, background field profiles in the strong field regime. The second is to go beyond the background field model itself, i.e.~the assumption that the background is not affected by back-reaction in the form of, for example, depletion.

In this paper we have put forward a framework in which back-reaction can be easily accounted for using a minor extension of the background field formalism; essentially the only change is that the background field is allowed to become complex, which corresponds to considering scattering between different initial and final coherent states.  In contrast to phenomenological models, though, unitarity is not violated. Further, our approach is exact in the sense that no approximation of the basic theory is needed, which may be an advantage regarding renormalisation~\cite{Epelbaum:2014yja}. 

Rules such as (\ref{in-red}) and (\ref{ut-red}) mean that additional (but lower order) diagrams must be calculated in order to completely describe depletion effects. However this is rather a (usually neglected) feature of strong field effects rather than a drawback of the method. Indeed diagrams which are usually dropped in ordinary perturbation theory can contain rich physics in background fields;  the \textit{disconnected} product of two tadpole diagrams, for example, contributes nontrivially to the two-loop two-point function in a background, as it can describe absorption by, and spontaneous emission from, the background. Such diagrams have been the focus of some attention recently, see~\cite{Gies:2016yaa} and~\cite{Edwards:2017bte,Ahmadiniaz:2017rrk} for related results.

We have given several elementary applications of our method which account for both weak and strong back-reaction effects. We have seen that depletion is described, at the level of the probability, by cross-terms between scattering amplitudes with different numbers of final state particles.  We have also introduced an extremisation principle in order to determine the `most likely' final coherent state profile given an initial state.

An interesting challenge for future is to extend our approach to background fields which are sourced; the immediate problem to overcome there is that the field will then have off-shell Fourier modes, so that it cannot be so simply extracted from the $S$-matrix using the displacement operators $D(z)$.

We can in future reinvestigate well-studied processes but including depletion. Consider for example multiple scattering events, such as photon emission or pair production; including depletion due to the first event will impact the second scattering event as the `background' field structure will have changed. This prompts us also to investigate the classical limit of our methods. How does the appearance of a complex potential relate to the dynamical change in a field interacting with a classical particle, as could be solved for using the classical Maxwell and Lorentz equations?

We have seen several features reminiscent of infra-red structures. As described above, scattering to a coherent state can be regarded as an inclusive process~\cite{Frantz}. It would be interesting to investigate the infra-red structure of our coherent state prescription in QED proper, with massless photons, especially in light of e.g.~\cite{Zwanziger:1973if} where complex backgrounds arise to account for the asymptotic dressing~\cite{Lavelle:1995ty} of particles due to infra-red effects.

While modern terrestrial experiments may not yet typically probe physical regimes in which beam depletion is significant, this situation will change~\cite{Kasper:2014uaa,Gelfer:2015ora}. The next step is therefore to apply the approach described above to phenomenological processes of interest. This will be pursued elsewhere.

\begin{acknowledgements}
A.I.~and D.S.~thank T.~Blackburn, S.~Bulanov, T.~Heinzl and M.~Marklund for useful discussions, and A.I.~thanks S.~Bragin, H.~Gies and H.~Ruhl for many useful comments. This project has received funding from the European Union's Horizon 2020 research and innovation programme under the Marie Sk\l odowska-Curie grant No.~701676 (A.I.), and by the UK
Science and Technology Facilities Council, Grant No. ST/G008248/1 (D.S.).
\end{acknowledgements}

\bibliography{depletion,lib_DS}

\end{document}